\documentclass[11pt]{article}
\usepackage{amsmath,amsfonts}
\usepackage{algorithmic}
\usepackage{array}
\usepackage{textcomp}
\usepackage{stfloats}
\usepackage{url}
\usepackage{verbatim}
\usepackage{graphicx}

\usepackage{xcolor}
\usepackage{caption}

\usepackage{subcaption}

\onecolumn
\begin{document}
	\title{Uncoordinated Cooperative OFDM Multi-Hop UAV Relay Networks Using Virtual Channels Based on All-Pass Filters 
	}
	\author{
		Noura Sellami \thanks{LETI Lab., ENIS, Univ. of Sfax, Tunisia, noura.sellami@enis.tn,} \and 
		Mohamed Siala \thanks{MEDIATRON Lab., SUP'COM, Univ. of Carthage, Tunisia, mohamed.siala@supcom.tn}
	}	
    \date{}
	\maketitle	
\begin{abstract}
In this paper, we propose an efficient transmission scheme for autonomous cooperative Orthogonal Frequency Division Multiplexing (OFDM) based multi-hop Unmanned Aerial Vehicle (UAV) relay networks. These systems often suffer from destructive interference at the destination node due to uncoordinated transmissions of common packets by cooperating UAVs. To address this issue, we introduce the concept of virtual transmit channels at each UAV, implemented using truncated all-pass filters (APFs). This approach ensures that all subcarriers benefit from comparable transmit powers, guaranteeing excellent performance when a single UAV is transmitting. In scenarios where multiple UAVs cooperate without coordination, the inherent randomness of the generated virtual channels facilitates cooperative diversity, effectively mitigating destructive interference. We further integrate this method with the distributed randomized space-time block coding (STBC) scheme to enhance transmission reliability. Additionally, we propose efficient algorithms for estimating the composite channels that combine both the true propagation channels and the virtual channels. Simulation results demonstrate that our proposed scheme significantly outperforms the classical phase dithering scheme across various scenarios.
\end{abstract}


\section{Introduction}

Unmanned Aerial Vehicle (UAV) relay networks have emerged as a promising technology to extend wireless coverage and provide efficient communication services in scenarios where terrestrial infrastructure is limited or unavailable, such as in disaster recovery, battlefield operations, and rural connectivity \cite{Mozaffari2019}. Thanks to their flexibility and rapid deployment capabilities, UAVs can be organized into multi-hop relay networks to support high-data-rate transmissions between distant source and destination nodes.

In cooperative UAV communication systems, UAVs assist one another by relaying messages across multiple hops. These systems can be broadly categorized into coordinated and uncoordinated architectures. In coordinated networks, relaying UAVs regularly exchange control information to jointly design and transmit their signals in a constructive manner. This coordination improves reliability but increases latency and leads to additional overhead in terms of energy and spectrum resources \cite{Woolsey2021}. In contrast, uncoordinated networks eliminate the need for explicit information exchange, enabling autonomous UAVs to operate independently. While this reduces overhead and simplifies deployment, it often results in destructive interference at the receiver due to lack of synchronization among transmitters \cite{Alouini2017}.

Recent studies have explored strategies to mitigate interference and enhance robustness in UAV-assisted communication systems. For instance, Reconfigurable Intelligent Surfaces (RIS) and Coordinated Multi-Point (CoMP) techniques have been employed to improve spectral efficiency in multi-UAV networks \cite{Chen2024}. Swarm-level coordination frameworks such as SwarmControl \cite{Bertizzolo2020} and interference-aware queuing strategies \cite{Ghazikor} have also been introduced to manage distributed transmissions. However, these solutions typically rely on centralized control or additional infrastructure, making them less suited to fully autonomous UAV operations.

\subsection*{Motivation and Research Gap}

In uncoordinated UAV relay networks, where UAVs may need to simultaneously forward the same data packet to enhance reliability, interference becomes a key limiting factor. Prior work has addressed this through two main approaches in single-carrier systems: Barrage Relay Networks (BRNs) \cite{Halford2010}, which apply random phase dithering to mitigate interference \cite{Hammerstrom04}, and randomized distributed Space-Time Block Coding (STBC) \cite{Scaglione2006}, where each single-antenna node sends an independent random linear combination of the codewords that would have been transmitted by the
elements of a multi-antenna system. Distributed STBC \cite{Scaglione2006} has been used in BRNs in \cite{Lee2023} and in UAVs networks in \cite{Bithas2023} over flat fading channels. The work of \cite{Scaglione2006} has been extended to the case of frequency selective channels in \cite{Sharp2006}, but only under the assumption of perfect channel knowledge.

Despite these advancements, there is still a lack of practical, low-complexity transmission frameworks that can operate effectively in uncoordinated, multi-hop UAV relay networks over frequency selective channels without requiring channel state information. This constitutes the main research gap addressed in this paper.

\subsection*{Objective and Solution Methodology}

The objective of this paper is to develop an interference-resilient, low-complexity transmission scheme tailored to uncoordinated UAV relay networks operating over frequency selective channels. Our methodology builds on Orthogonal Frequency Division Multiplexing (OFDM), a key technology in 4G, 5G, and emerging 6G networks \cite{Zhang2019}. Each UAV independently generates a set of virtual transmit channels that alter the transmitted signals in a way that prevents destructive interference at the receiver.

\subsection*{Contributions}

The main contributions of the paper are as follows:
\begin{itemize}
    \item \textbf{Virtual channel generation:} We introduce a method based on truncated all-pass filters (APFs) \cite{Manolakis2011} to generate virtual channels that preserve power across subcarriers. This ensures reliable transmission when only one UAV is active, and prevents destructive interference when multiple UAVs transmit simultaneously without coordination.
    
    \item \textbf{Integration with randomized STBC:} The generated virtual channels serve as randomization vectors in a distributed randomized STBC framework \cite{Scaglione2006}, enabling spatial diversity without coordination.

    \item \textbf{Efficient channel estimation techniques:} We propose two low-complexity algorithms for estimating the composite channels (true channels plus virtual channels): (i) LS frequency domain channel estimation with time domain denoising and (ii) cyclic delay based channel estimation.
\end{itemize}

This framework addresses the core challenges of destructive interference and channel uncertainty in uncoordinated multi-hop UAV communications, offering a practical and scalable solution for future UAV networks.

The remainder of this paper is organized as follows. Section 2 presents the system model. Section 3 describes our method for generating virtual channels. In Section 4, we propose efficient methods for channel estimation. Section 5 provides the simulation results. Section 6 concludes the paper. Throughout this paper, vectors are underlined, and matrices are written in bold, either in lower or upper case. Additionally, $(.)^*$, $(.)^T$, and $(.)^H$ denote conjugation, transposition, and Hermitian transposition, respectively, and $\mathbf{I}_K$ represents the $K \times K$ identity matrix.

	\section{System model and proposed diversity scheme}
\begin{figure}[tb]
	\centerline{\includegraphics[height=5cm,width=9cm]{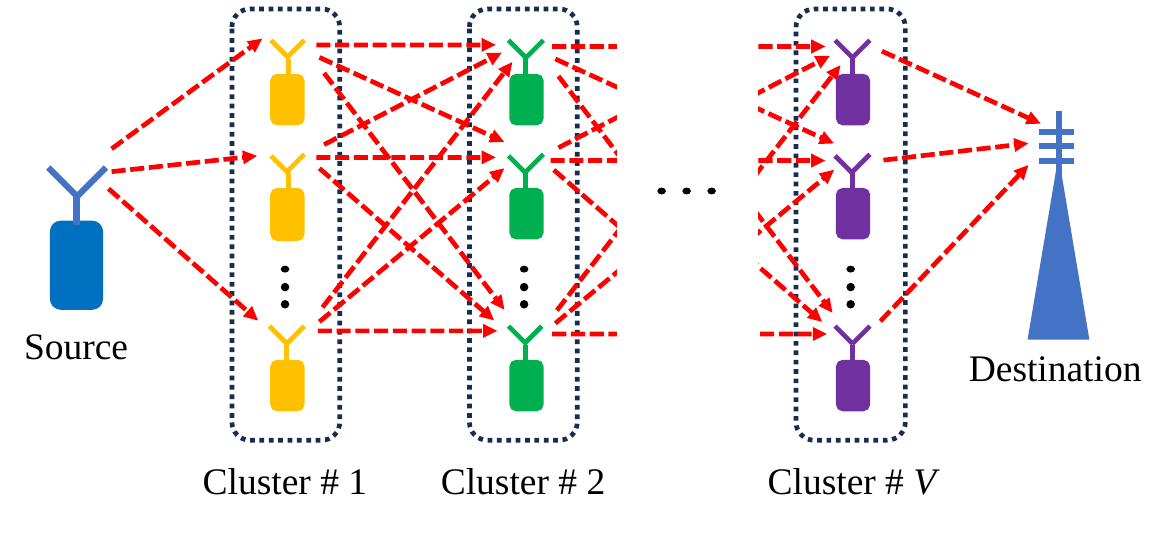}}
	\caption{Multi-hop cluster based UAV relay network }
	\label{fig1}
\end{figure}
As shown in Figure 1, we consider a wireless multi-hop UAV relay network where the source transmits its data to the destination via multiple UAVs locating in $V$ intermediate clusters. Each cluster contains UAVs that receive common information sequences from the UAVs of the preceding cluster (or from the source if they are in the first cluster) and aim to transmit them to the UAVs of the following cluster (or to the destination if they are in the $V^{\text {th}}$ cluster) without any coordination between each other. We consider that the UAVs remain stationary while providing the service. This assumption reflects practical deployment scenarios such as aerial monitoring, fixed-position surveillance, or temporary hotspot coverage, where UAVs hover in place to ensure reliable communication. We assume that the nodes involved during the relaying stage in the network use a basic single transmit and receive antenna. Thus, no explicit transmit or receive diversity is available in the network. We consider transmissions over frequency selective channels, and we assume that the transmitted signals are based on OFDM systems and organized as in WiFi systems \cite{Alouini2017}.

We consider in this work the sampled mode. We assume that the transmissions are organized into blocks of $T$ OFDM symbols. Let $K$ be the number of subcarriers per OFDM symbol. The first $T_p$ OFDM symbols are pilot symbols constituting the Long Training Sequence (LTS). Without loss of generality, we consider the transmission during one hop from $U$ UAVs of a cluster to one receiver (an UAV of the following cluster or the destination). We assume that the information messages are perfectly known at the $U$ UAVs. We consider that the channel between the $u^{\text {th}}$ UAV, for $0\leq u \leq U-1$, and the receiver is invariant during a block of $T$ OFDM symbols and changes independently from block to block. It is modeled as a Rice fading channel with $L$ taps \cite{Goddemeier}. Let $\underline{h}_{u}$ be the $L\times 1$ channel impulse response and $\underline{H}_{u}=(H_{u}(0),\ldots,H_{u}(K-1))^T$ be its Fast Fourier Transform (FFT). 

As shown in Figure 2, we consider a Bit Interleaved Coded Modulation (BICM) scheme at each UAV \cite{Sellami2002}. The data sequence (common to all UAVs) is encoded, at each UAV, by a convolutional channel encoder. Then, the coded sequence is interleaved by an interleaver $\pi$ and mapped to the same modulated sequence. We propose in the following to make each cooperating transmitting UAV play with its unique antenna the role of a transmitting antenna array in a standard Multiple Input Single Output (MISO) system using a distributed STBC scheme. For the sake of clarity, we will consider the transmission of one OFDM symbol, $\underline{S}=(S_0,\ldots,S_{K-1})$. We assume that the $K$ subcarriers are organized into vectors of $P$ subcarriers each, where $K$ is a multiple of $P$. At each UAV, for $0 \leq k \leq K-1$ and $k$ multiple of $P$, the vector $(S_{k},S_{k+1},...,S_{k+P-1})$ is mapped to the $P\times N$ matrix $\mathbf{S}(k)$, as is done in standard orthogonal STBC, where $N$ is the number of antennas in the underlying code and $N \leq P$ \cite{Sharp2006}.
For example, when $P=N=2$, the STBC is the Alamouti code \cite{Alamouti1998} and the code matrices, for $0 \leq k \leq K-1$ and $k$ multiple of 2, are,
\begin{equation}
\mathbf{S}(k)=\begin{bmatrix} S_{k} & S_{k+1} \\ -S_{k+1}^* & S_{k}^* \end{bmatrix}.
\end{equation}
We propose, to achieve complete transmit diversity, that each UAV $u$ generates $N$ virtual channels (randomization vectors) of length $K$ leading to a $N\times K$ virtual transmit channel matrix $\mathbf{R_u}$ whose entry in row $n$ and column $k$ is denoted by $R_u(n,k)$. We assume in the following that $k$ is a multiple of $P$. The transmitted vector at carriers $k, k+1,..., k+P-1$ is $\mathbf{S}(k)\underline{R}_{u}(k)$ where $\underline{R}_{u}(k)$ is the $k^{\text {th}}$ column of $\mathbf{R_u}$. Notice that the vectors $\underline{R}_{u}(i)$ where $i$ is not a multiple of $P$ will not be used in our system. The randomized STBC can then be expressed as a double mapping:
\begin{equation}\label{mapping}
( S_{k},S_{k+1},...,S_{k+P-1} ) \rightarrow \mathbf{S}(k)\rightarrow \mathbf{S}(k)\underline{R}_{u}(k).
\end{equation}
After aggregation of the vectors obtained in (\ref{mapping}), Inverse FFT (IFFT) and Cyclic Prefix (CP) addition are applied. The length of the CP is $T_{cp}$ samples where $T_{cp}\geq L-1$. 
The signal is then transmitted over the channel. At the receiver, the CP is removed and FFT is performed. Let $\underline{H}^{eq}(k)= \left( H^{eq}(0,k),...,H^{eq}(N-1,k)\right)^T $, be the frequency response of the composite MISO channel, which incorporates the true propagation channels and the virtual channels, at subcarrier $k$ where $H^{eq}(n,k)=\sum_{u=0}^{U-1} R_u(n,k)H_{u}(k)$. We assume that $H^{eq}(n,k)\approx H^{eq}(n,k+p)$ for $k$ multiple of $P$ and $1\leq p\leq P-1$. Thus, at the receiver, the received vector at carriers $k, k+1,..., k+P-1$, $k$ being an integer multiple of $P$, can be approximated by,	 
\begin{equation}\label{rec}
\underline{Y}(k) \approx \mathbf{S}(k)\underline{H}^{eq}(k)+\underline{W}(k)
\end{equation}
where $\underline{W}(k)$ is complex additive white Gaussian noise
(AWGN) vector with zero mean and variance $N_0$.
\\ At the receiver, the composite channel is estimated, as will be explained in Section 4. Then, STBC decoding with soft outputs is performed. The outputs are demodulated, deinterleaved, and decoded by a soft input Viterbi decoder.
In the following, we present our method to generate virtual channels.
\begin{figure}[tb]
	\centerline{\includegraphics[height=7cm,width=16cm]{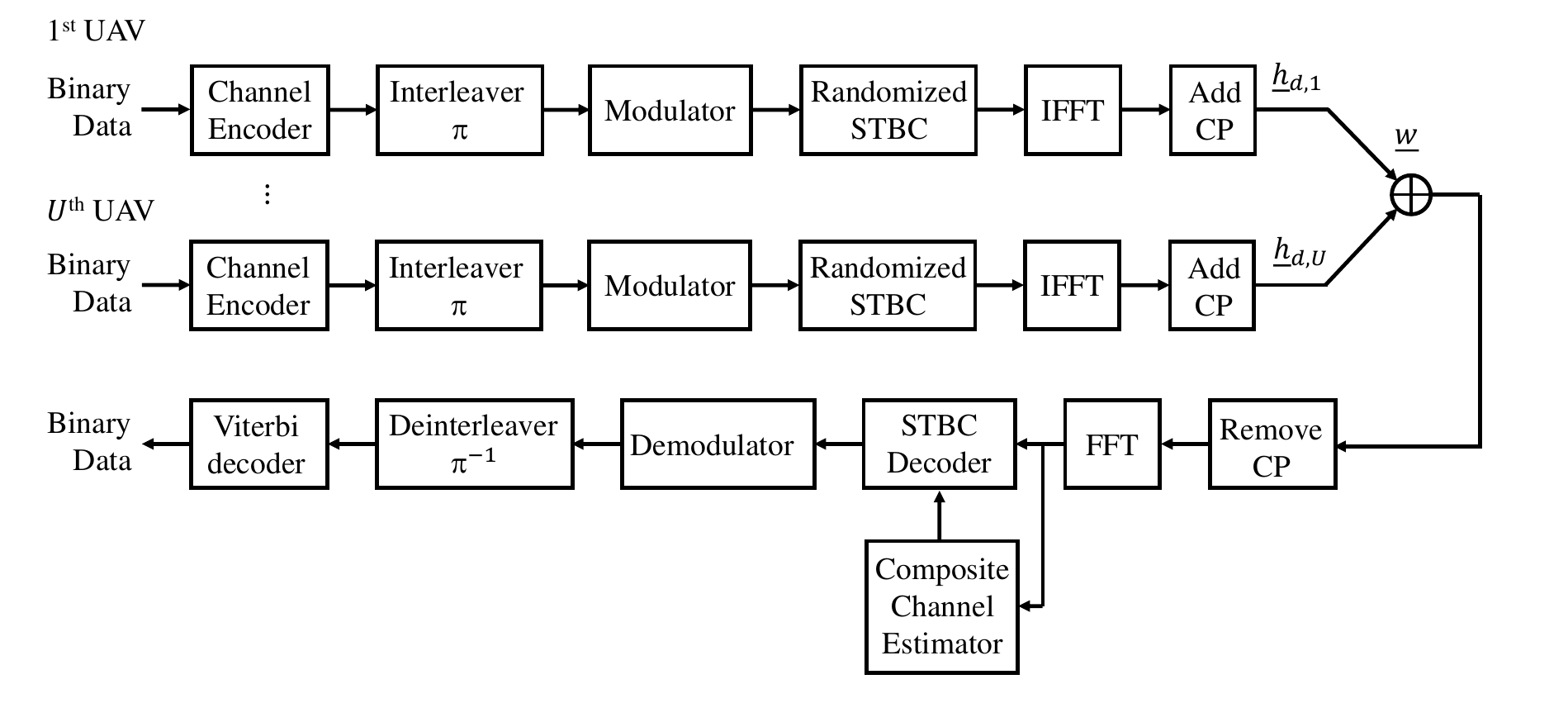}}
	\caption{Transmitters and receiver structures}
	\label{fig2}
\end{figure}	
\section{Generation of virtual channels}
One simple method to draw virtual channels is to consider random non-time dispersive channels for each transmitting node. This approach could potentially lead to the destructive addition of replicas of the signal coming from different UAVs. In order to bring additional randomness, another method is to use multipath channels. The drawback of this approach is the inherent frequency selectivity of the underlying channels, which consists of allocating different powers to the different subcarriers of the OFDM system and leads to performance degradation when only one UAV is transmitting. To alleviate this phenomenon and make all subcarriers benefit from comparable transmit powers while still preserving most of the benefits of multipath randomness, we generate virtual channels as APFs since they have a flat magnitude frequency response \cite{Manolakis2011}. 
The transfer function of an APF has all poles and zeros occurring in conjugate reciprocal pairs. An APF of order $M$ can be conceived as a cascade of $M$ first order APFs. Its transfer function takes the form,
\begin{equation}
G(z)=\prod_{m=1}^M G_{m}(z)=\prod_{m=1}^M \frac{p_m^*-z^{-1}}{1-p_mz^{-1}},
\end{equation}
where $G_{m}(z)$ is the transfer function of the $m^{\text {th}}$ first order APF and $p_m$ is its pole. 
We consider here causal filters. Then, the impulse response of the $m^{\text {th}}$ first-order APF is infinite with the $i^{\text {th}}$ tap given by, 
\begin{equation}
\begin{array}{ll}
g_{i,m}=-p_m^*, \text{for } i=0 \\ 
g_{i,m}=p_m^i (\frac{1}{p_m}-p_m^*), \text{for} i>0.
\end{array}
\end{equation}
Let $g_i$, for $i\geq 0$, be the $i^{\text {th}}$ tap of the impulse response of the APF of order $M$. In order to allow the generation of the virtual channels and the estimation of impulse responses of the composite channels, we consider truncated versions of the APFs. Thus, we generate the taps of the $n^{\text {th}}$ temporal virtual channel for the $u^{\text {th}}$ UAV as $r_{u}(n,i)=g_i$ for $0 \le i \leq T_c-1$ and $r_{u}(n,i)=0$ elsewhere. Then, $\left( R_{u}(n,0),...,R_{u}(n,K-1)\right)^T $ is given by the FFT of $\left( r_{u}(n,0),...,r_{u}(n,T_c-1)\right)^T $. Notice that we generate a different APF for each virtual channel of each UAV. As $T_c$ decreases, the property of having a flat magnitude response is less and less verified. Thus, a compromise has to be found between this property and the accuracy of the channel estimation. The choice of $T_c$ as well as that of the poles of the APFs and the order $M$ will be discussed in Section 5. We present, in the following, two methods for estimating the composite channels.

\section{Channel estimation}
We assume that all UAVs use the same LTS. In the first method, the training sequence covers distinct subcarriers, typically regularly spaced, offering a subsampled frequency estimation of each composite channel corresponding to each dimension of the STBC. In the second one, distinct cyclic time shifts of the training sequence are used for each composite channel. 
In the following, we denote by $\underline{h}_n^{eq}$ the temporal response of the composite channel corresponding to the virtual channel $n$, for $0\leq n\leq N-1$, given by the IFFT of $\left( H^{eq}(n,0),...,H^{eq}(n,K-1)\right)^T $. We present the channel estimation methods for one OFDM pilot symbol, $\underline{P}=(P_0,\ldots,P_{K-1})$.
\subsection{LS frequency domain channel estimation with time domain denoising}
We assume that $k$ is a multiple of $P$ and $0\leq k \leq K-1$. At carrier $k+n$, for $0\leq n\leq N-1$, each UAV $u$ transmits $P_{k+n}R_u(n,k)$, where $P_{k+n}$ are the pilot tones. Notice here that if $N <P$, some pilot tones are not used, and therefore not transmitted. Then, IFFT and CP insertion are applied. Figure 3.a shows the design of the preamble of user $u$ before transmission when the LS frequency domain channel estimation is used, with $N=P=2$. After transmission over the channels, FFT and removal of the CP, the received signals at carriers $k+n$, for $0 \leq n \leq N-1$, can be approximated as,
\begin{equation}
Y_{k+n}\approx P_{k+n}H^{eq}(n,k)+W_{k+n},
\end{equation}
where $W_{k+n}$ are the noise samples.
Thus, the LS estimations of the composite channels taps $H^{eq}(n,k)$, where $k$ is a multiple of $P$, are,
\begin{equation}
\begin{array}{ll}
\hat{H}^{eq}_{LS}(n,k)=\frac{Y_{k+n}}{P_{k+n}}.
\end{array}
\label{LS}
\end{equation}
Since we do not obtain a complete estimation of the channel transfer function for each composite channel, we propose to recover it by interpolation for all subcarriers. We then apply an IFFT to the estimated frequency channel to obtain an estimation $\underline{\hat{h}}^{eq}_n$ of the composite channel impulse response $\underline{h}^{eq}_n$, for $0\leq n \leq N-1$. We recall that each composite channel $\underline{h}^{eq}_n$ is equal to the sum of the convolution products of the virtual channels and the true propagation channels between the UAVs and the receiver, and then its length is equal to $T_c+L-1$. Thus, the estimated taps $\hat{h}^{eq}_n(i)$ are forced to zero for $i>T_c+T_{cp}$.

While minimum mean square error (MMSE) estimation generally yields better performance, it requires prior knowledge of noise variance and channel statistics. In contrast, LS estimation, combined with denoising, offers a favorable trade-off between complexity and robustness, particularly in decentralized UAV scenarios where such statistical information is difficult to acquire.

\subsection{Cyclic delay channel estimation}
To avoid interpolation, we perform in this second method cyclic time shifts of the LTS of $\frac{Kn}{N}$, for $0\leq n \leq N-1$ for each dimension $n$ of the STBC. We assume here that $K$ is a multiple of $N$  to ensure that these cyclic time shifts result in integer sample shifts. Thus, we consider, for $0\leq k \leq K-1$, 
\begin{equation}
P_k^{(n)}=P_k \exp\left(\frac{-2j\pi kn}{N}\right). 
\end{equation}
It is equivalent to having $N$ different OFDM pilot symbols, one for each virtual channel $n$ with components $P_k^{(n)}$, for $0\leq n \leq N-1$ and $0\leq k \leq K-1$.

Each UAV $u$ generates the signal to be transmitted at subcarrier $k$ as $\sum_{n=0}^{N-1} P_k^{(n)}R_u(n,k)$. Then, IFFT and CP addition are applied. 
At the receiver, after CP removal and FFT, the received signal is $\underline{\tilde{Y}}=\left( \tilde{Y}_0,..,\tilde{Y}_{K-1} \right)^T $. The channel estimator calculates, for $0 \leq k \leq K-1$,
\begin{equation}
\tilde{H}^{eq}_k=\frac{\tilde{Y}_k}{P_k}.
\end{equation}
The IFFT is applied to $\underline{\tilde{H}}^{eq}=\left(\tilde{H}^{eq}_0,..,\tilde{H}^{eq}_{K-1} \right)^T$ leading to
$\underline{\tilde{h}}^{eq}=\left( (\underline{ \tilde{h}}^{eq}_0)^T,(\underline{\tilde{h}}^{eq}_1)^T, \ldots ,(\underline{\tilde{h}}^{eq}_{N-1})^T \right)^T $ where
$\underline{\tilde{h}}^{eq}_n=\left(\tilde{h}^{eq}_{\frac{nK}{N}},\tilde{h}^{eq}_{\frac{nK}{N}+1},...,\tilde{h}^{eq}_{\frac{(n+1)K}{N}-1}\right)^T $ is the shifted estimate of the composite channel $\underline{h}^{eq}_n$, for $0 \leq n \leq N-1$. Since the maximum length of the impulse response of the composite channel is $T_c+T_{cp}$, the estimate of $\underline{h}^{eq}_n$ is given by $\left(\tilde{h}^{eq}_{\frac{nK}{N}},\tilde{h}^{eq}_{\frac{nK}{N}+1},...,\tilde{h}^{eq}_{\frac{nK}{N}+T_c+T_{cp}-1}\right)^T $. Notice that in this case, the footprints of the composite multipath channels corresponding to the cyclic time shifts of each STBC code dimension should not overlap in time. Thus, we assume, when cyclic delay channel estimation is used, that $T_c\leq \frac{K}{N}-T_{cp}$.

In the following, we provide a complexity analysis of both methods to assess their feasibility for real-time implementation in UAV relay networks.
\subsection{Complexity Analysis}

The LS channel estimation requires $\mathcal{O}(\frac{KP}{N})$ multiplications for pilot based transmission, with an additional $\mathcal{O}(K)$ operations for interpolation. The dominant computational cost in this approach comes from the IFFT, which has a complexity of $\mathcal{O}(K \log K)$. In the cyclic delay approach, $O(NK)$ multiplications are required due to the application of cyclic time shifts. The IFFT again dominates the total complexity with $\mathcal{O}(K \log K)$. In summary, both methods have a dominant complexity of $\mathcal{O}(K \log K)$. The cyclic delay method introduces additional complexity due to the cyclic time shifts, but it avoids the need for interpolation, which may be advantageous depending on the system constraints.


		\section{Simulation results}

We conduct our simulations based on the system model described in Section 2. We use MATLAB for the implementation, adhering strictly to the parameters and assumptions specified in the system model. The network consists of $U$ UAVs transmitting the same data to a single receiver. To evaluate the impact of cooperation, we simulate different numbers of UAVs $U$. We consider the transmission of blocks of $T=20$ OFDM symbols with $T_p=2$ pilot OFDM symbols forming the LTS. The length of the cyclic prefix is $T_{cp}=16$ samples. We assume that $K=128$ subcarriers are used for each OFDM symbol. The input bit sequence in each UAV is first encoded with a non-recursive non-systematic convolutional encoder with rate $\frac{1}{2}$ and generator polynomials $(7,5)$. We use the Quadrature Phase Shift Keying (QPSK) modulation. Each channel between a UAV and the receiver is modeled as a Rician fading channel with $L=3$ taps and an exponential power delay profile. We consider two values of the Rician factor, $K_{\text{Rice}} = 10$~dB and $20$~dB, to model moderate and strong line of sight (LoS) conditions, respectively. The Alamouti coding is considered. Thus, $P=N=2$. We performed simulations to compare LS channel estimation with denoising and cyclic delay based estimation. The results showed negligible differences in the considered slow-fading scenarios. Therefore, for clarity and brevity, we present only the results obtained using cyclic delay estimation in the following.

\begin{figure}[tb]
    \centering
    \begin{subfigure}{0.49\columnwidth} 
        \centering
        \includegraphics[width=1\textwidth]{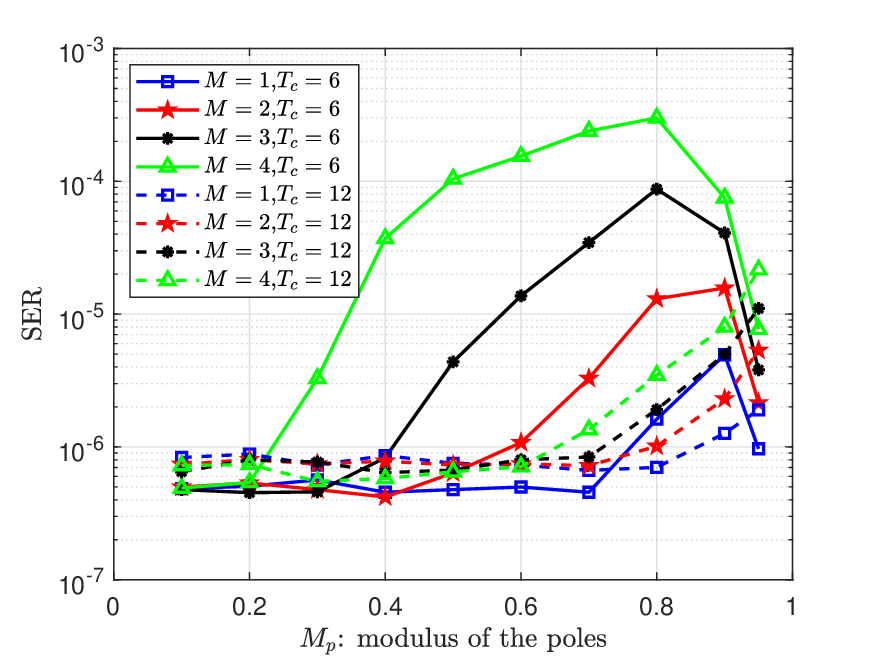}
        \captionsetup{labelformat=empty} 
        \caption{(a)}
        \label{fig3a}
    \end{subfigure}
    \hfill
    \begin{subfigure}{0.49\columnwidth}
        \centering
        \includegraphics[width=1\textwidth]{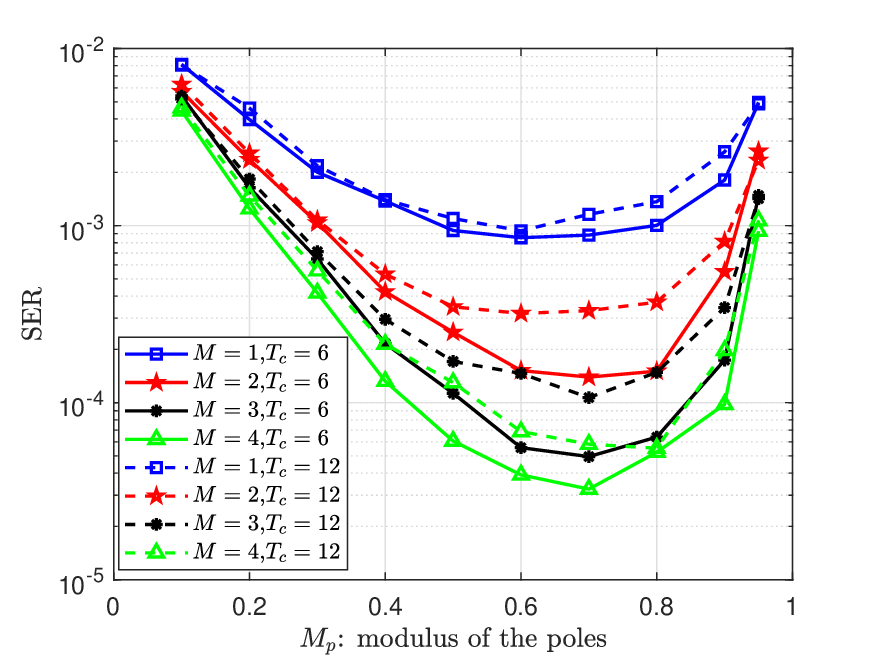}
        \captionsetup{labelformat=empty} 
        \caption{(b)}
        \label{fig3b}
    \end{subfigure}
    
    \caption{SER versus the modulus of the poles  for different values of $M$ when the truncated APF virtual channels are used with $T_c=6$ (solid curves) $T_c=12$ (dashed curves), with cyclic delay channel estimation: (a) $U=1$, (b) $U=2$.}
    \label{fig4}
\end{figure}

We first discuss the choice of the parameters involved in the generation of the truncated APFs. For the sake of simplicity, we assume that the moduli of all poles are all equal to $M_p<1$. Thus, the poles are randomly distributed on the circle of radius $M_p$. Figs. \ref{fig3a} and \ref{fig3b} show the Symbol Error Rate (SER) curves versus $M_p$, for $K_{\text{Rice}}=20$ dB, for $U=1$ and $U=2$ respectively, for $M\in\{1,2,3,4\}$ and $T_c \in \{6,12\}$. Here, we set $\frac{E_s}{N_0}=30$ dB, where $E_s$ is the energy per QPSK symbol. Fig. \ref{fig3a} shows that the use of poles with modulus 0.1 leads to the best performance overall. When $T_c=6$, the performance degrades as $M_p$ increases and then improves. This is due to the nature of the impulse response of the APF. Actually, the function giving the modulus of the $i^{\text {th}}$ tap of an APF versus $M_p$ is concave. Thus, as the poles become closer to zero or to one, the modulus of the taps of the APF decreases rapidly, and then the truncation has less effect, and the gains of the truncated APFs on all subcarriers remain close to one. When only one UAV is transmitting, this property is important to guarantee almost the same transmitted powers on all subcarriers. We also notice that the truncature has more effect when the order of the APFs increases.
  Fig. \ref{fig3b} shows that when $U=2$, the use of poles with a modulus roughly equal to 0.7 is better than the other choices. Indeed, when many UAVs are transmitting, each composite channel is the combination of the different UAVs channels and thus the property of equal transmitted powers on the different subcarriers is lost. Contrary to the case where $U=1$, the increase of $M$ leads to better performance. Indeed, when $M$ increases, the phase of the frequency response of the APF becomes more selective and varies faster, which brings more diversity. Notice that when $U\geq 3$, the obtained results are similar to those obtained when $U=2$. 
  
  In the following simulations, to achieve a good compromise between the performance obtained when $U=1$ and that obtained when $U$ is higher, we set $M=4$, $T_c=12$ and $M_p=0.7$. In order to provide further insight into the impact of virtual channel generation using APFs, Fig. 4 shows the magnitude of the composite channel frequency response, $|H^{eq}(0,k)|$, as a function of the subcarrier index $k$, for $K_{\text{Rice}} = 20$~dB and $U\in \{1,2,3,5,7\}$. When $U=1$, we observe that $|H^{eq}(0,k)| \approx 1$, which ensures that all subcarriers benefit from comparable transmit powers. As $U$ increases, the channels become more frequency selective, resulting in improved diversity gain. Naturally, the total transmitted power also increases with $U$.
\begin{figure}[tb]
		\centerline{\includegraphics[width=11cm]{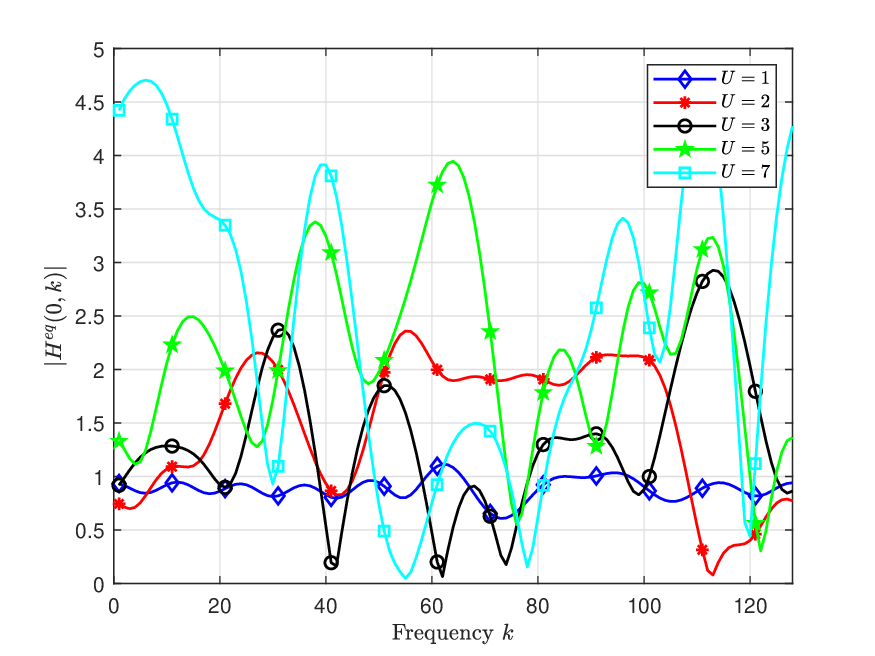}}
		\caption{Magnitude of the composite channel frequency response $|H^{eq}(0,k)|$ as a function of the subcarrier index $k$.}
		\label{fig4}
	\end{figure}
 We propose now to compare our method with a generalization of phase dithering used in BRNs \cite{Halford2010,Hammerstrom04} to our case. Notice that phase dithering is equivalent to the uniform phase randomization proposed in \cite{Scaglione2006}. The generalization of phase dithering consists in generating the frequency virtual channel coefficients as complex exponentials, whose phases are constant on a group of $P$ subcarriers and change independently from group to group (to respect the STBC constraints). The phases are uniformly distributed on $[ 0,2\pi [ $. We name this method space time (ST) block phase dithering. Since the composite virtual channels are not defined in the temporal domain, channel estimation can be done only in the frequency domain in this case.
\begin{figure}[tb]
    \centering
    \begin{subfigure}{0.49\columnwidth} 
        \centering
        \includegraphics[width=1\textwidth]{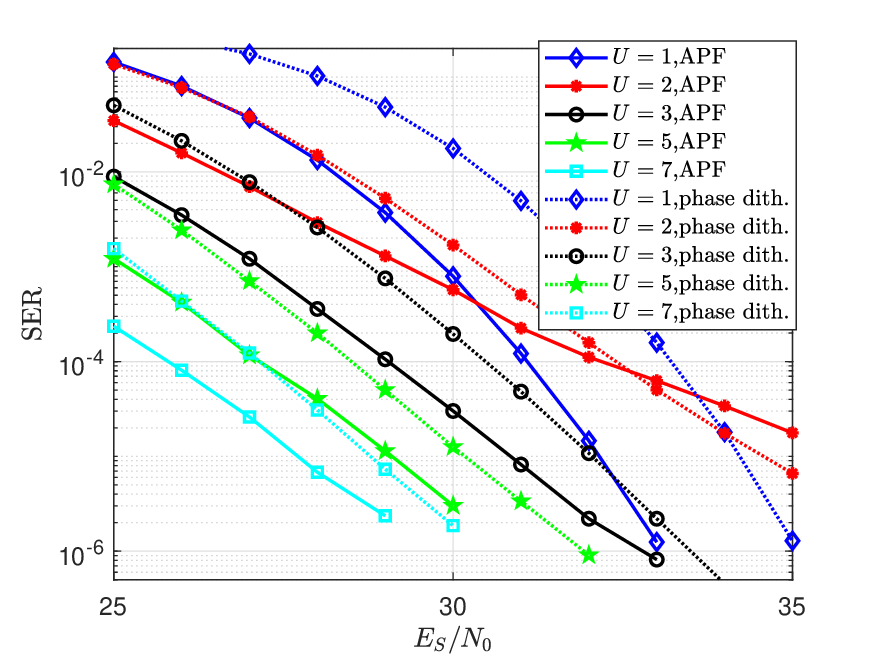}
        \captionsetup{labelformat=empty} 
        \caption{(a)}
        \label{fig5a}
    \end{subfigure}
    \hfill
    \begin{subfigure}{0.49\columnwidth}
        \centering
        \includegraphics[width=1\textwidth]{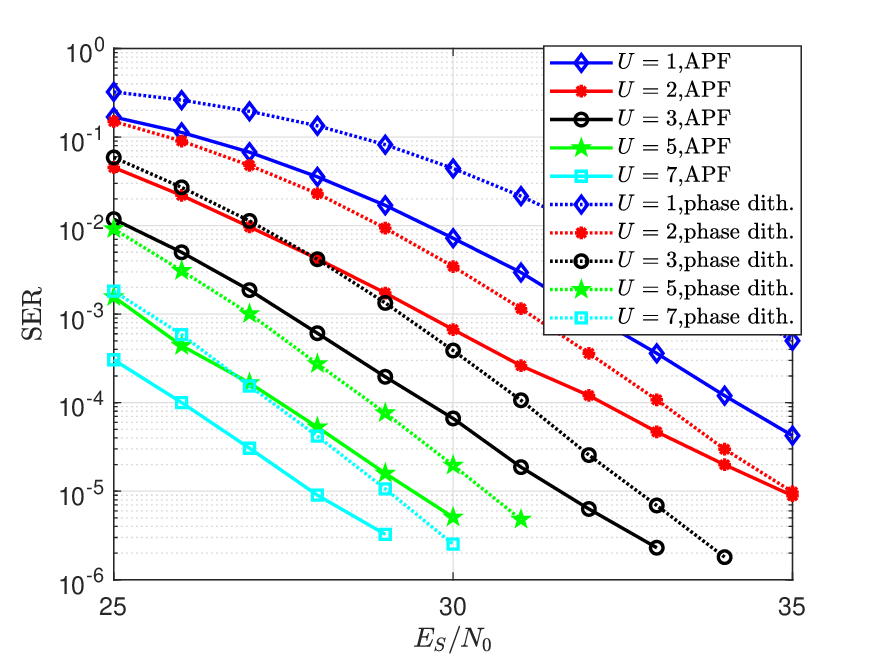}
        \captionsetup{labelformat=empty} 
        \caption{(b)}
        \label{fig5b}
    \end{subfigure}
    
    \caption{SER versus $E_s/N_0$  for different values of $U$ when the truncated APFs are used with $M=4$, $T_c=12$ and cyclic delay channel estimation (solid curves) and when the ST block phase dithering is used with LS frequency channel estimation (dotted curves): (a) $K_{\text{Rice}}=20$ dB, (b) $K_{\text{Rice}}=10$ dB.}
    \label{fig4}
\end{figure}
   Figs. \ref{fig5a} and \ref{fig5b} show the SER curves versus $\frac{E_s}{N_0}$, when virtual channels are generated as truncated APFs of order $M=4$ (solid curves) and using ST block phase dithering (dotted curves), for $U\in \{1,2,3,5,7\}$, $K_{\text{Rice}}= 20$ dB and $K_{\text{Rice}}= 10$ dB respectively. The length of the impulse responses of the truncated APFs is $T_c=12$. When the truncated APFs are considered, the composite channels are estimated using the cyclic delay channel estimation. When the ST block phase dithering is considered, we use the LS frequency domain channel estimation (\ref{LS}) which is the only choice in this case since the virtual channels are not defined in the temporal domain. This inevitably degrades the performance due to noisy estimates on the $K$ subcarriers.
   When $U=1$, the truncated APFs and the ST block phase dithering achieve almost the same diversity orders. The curves of SER have the same slopes for both schemes. However, our method based on truncated APFs leads to a significant performance gain compared to ST block phase dithering due to the better estimates of the channels it uses. 
  	When $U \geq 2$, our method achieves a lower diversity order compared to phase dithering since for the latter the phase of each composite channel is very selective. Thus, the curves of SER obtained by using our method when $U=2$ have lower slopes. As $U$ increases, the diversity orders achieved by both methods become closer, even if our scheme still achieves a lower diversity. Thus, when $U=2$, due to the different slopes and the gap caused by the difference in the qualities of channel estimates, the curves intersect when $\frac{E_s}{N_0}=32.5$ dB for $K_{\text{Rice}}= 20$ dB and when $\frac{E_s}{N_0} \geq 35$ dB for $K_{\text{Rice}}= 10$ dB. 
   	When $U\geq 3$, since the slopes become closer, the intersections of the curves do not appear on the figures for the considered range of $\frac{E_s}{N_0}$. Instead, the intersections would occur at higher values of $\frac{E_s}{N_0}$. Notice that the narrowing of the performance gap between both methods, when $U$ increases, can be attributed to the increased total transmitted power, which enhances the accuracy of channel estimation. Notably, this effect is particularly beneficial for phase dithering, where channel estimation is performed in the frequency domain. In summary, our method consistently outperforms phase dithering in practical scenarios where channel estimation is required. While phase dithering may achieve better performance in the idealized case of near-perfect channel estimation at very high \( \frac{E_s}{N_0} \), this assumption is unrealistic. In practical conditions, the reliance on frequency domain channel estimation significantly degrades the performance of phase dithering, reinforcing the advantage of our approach.

 	\section{Conclusion}
In this paper, we proposed an efficient transmission scheme based on virtual transmit channels for uncoordinated OFDM based multi-hop UAV relay networks. We generated these channels using truncated APFs to make all subcarriers benefit from comparable transmit powers and guarantee excellent performance when one UAV is transmitting. The randomness of the virtual channels prevents destructive interference at the receiver when multiple UAVs cooperate in an uncoordinated manner. We also integrated our scheme with a distributed, randomized STBC. We proposed efficient algorithms to estimate composite channels that combine true propagation channels and virtual channels. Simulation results showed that our proposed scheme relying on APFs, significantly outperforms, in practical scenarios, the one using phase dithering. While our current model is based on stationary UAVs with frequency selective channels, future work will address the challenges posed by dynamic environments, where rapidly moving UAVs introduce channels time selectivity. We also plan to explore more complex network topologies in future research. 
Moreover, we acknowledge the existence of alternative techniques for channel estimation in the literature, such as pilot optimization   and deep learning based methods \cite{Soltani}. A comparative study with such methods remains an interesting direction for future work.

  \section*{Declarations}  

  \begin{itemize}
  	\item Funding: No external funding was received to support this research.
  	\item Availability of data and material: The data and material used in this paper were developed by the authors and are not made publicly available.
  \end{itemize}

\end{document}